\documentclass[12pt]{article}

\usepackage[utf8x]{inputenc}
\usepackage{xcolor, graphicx}

\usepackage{hyperref}
\hypersetup{%
	colorlinks=true,
	breaklinks=true,
	linkcolor=red,
	anchorcolor=black,
	citecolor=brown,
	filecolor=blue,
	menucolor=red,
	urlcolor=blue,
}

\usepackage[square,numbers]{natbib}
\usepackage{mathtools, amsfonts, amssymb}
\usepackage{lmodern}
\numberwithin{equation}{section}
\mathtoolsset{showonlyrefs}

\usepackage{enumerate}
\usepackage{comment}
\usepackage{subcaption}
\usepackage{cleveref}
\usepackage{siunitx}
\usepackage{tikz}
\usetikzlibrary{positioning, shapes.misc}

\usepackage{numprint}
\usepackage{booktabs}

\usepackage{chngcntr}
\usepackage{apptools}
\AtAppendix{\counterwithin{lemma}{section}}

\DeclareSIUnit{\octet}{o}
\newcounter{thm}[section]
\newcounter{appen}[section]

\newcounter{rem}
\newtheorem{remark}[rem]{Remark}

\newcommand{\EE}{\mathbb{E}}

\newcommand{\RR}{\mathbb{R}}

\newcommand{\TT}{\mathcal{T}}
\newcommand{\HH}{\mathcal{H}}

\newcommand{\sLp}[2]{\mathcal{L}^{#1}(#2)}
\newcommand{\inLp}[1]{\langle#1\rangle}

\newcommand{\inH}[1]{\langle\!\langle#1\rangle\!\rangle}

\newcommand{\Xnp}{X_n^{(p)}} 
\newcommand{\Xp}[1]{X^{(#1)}} 
\newcommand{\mup}[1]{\mu^{(#1)}} 

\newcommand{\pointt}{\boldsymbol{t}}
\newcommand{\points}{\boldsymbol{s}}

\title{\textsf{FDApy}: a Python package for functional data}
\author{%
Steven Golovkine\thanks{MACSI, Department of Mathematics and Statistics, University of Limerick, Limerick, Ireland, \href{mailto:steven.golovkine@ul.ie}{steven.golovkine@ul.ie}}
}
\date{\today}

\begin{document}

\maketitle

\begin{abstract}
We introduce \textsf{FDApy}, an open-source Python package for the analysis of functional data. The package provides tools for the representation of (multivariate) functional data defined on different dimensional domains and for functional data that is irregularly sampled. Additionally, dimension reduction techniques are implemented for multivariate and/or multidimensional functional data that are regularly or irregularly sampled. A toolbox for generating functional datasets is also provided. The documentation includes installation and usage instructions, examples on simulated and real datasets and a complete description of the API. \textsf{FDApy} is released under the MIT license. The code and documentation are available at \url{https://github.com/StevenGolovkine/FDApy}.
\end{abstract}


\section{Introduction} 
\label{sec:introduction}

Functional data analysis (FDA) is a statistical methodology for analyzing data that can be characterized as functions. These functions could represent measurements taken over time, space, frequency, probability, etc. The goal of FDA is to extract meaningful information from these functions and to model their behavior. See, e.g., \cite{ramsayFunctionalDataAnalysis2005,horvathInferenceFunctionalData2012a,kokoszkaIntroductionFunctionalData2017} for some references on FDA. FDA has been successfully applied in different contexts, such as identifying patterns of movements in sport biomechanics \cite{warmenhovenBivariateFunctionalPrincipal2019}, analyzing changes in brain activity in neuroscience \cite{songSparseMultivariateFunctional2022}, fault detection of batch processes \cite{wangFaultDetectionBatch2015} or in autonomous driving \cite{golovkineClusteringMultivariateFunctional2022}.

In order to apply FDA to real datasets, there is a need for appropriate softwares with up-to-date methodological implementation and easy addition of new theoretical developments. Currently, most FDA softwares are implemented in R \cite{rcoreteamLanguageEnvironmentStatistical2021}. The seminal R package for FDA is \textsf{fda} \cite{ramsayFdaFunctionalData2023}, based on foundational work by \cite{ramsayFunctionalDataAnalysis2005} and \cite{ramsayFunctionalDataAnalysis2009}. Many specialized R packages for FDA are built upon \textsf{fda}. For instance, \textsf{FDboost} \cite{brockhausBoostingFunctionalRegression2020} and \textsf{refund} \cite{goldsmithRefundRegressionFunctional2023} are used for regression and classification, while \textsf{funFEM} \cite{bouveyronFunFEMClusteringDiscriminative2021} and \textsf{funLBM} \cite{bouveyronFunLBMModelBasedCoClustering2022} are tailored for clustering. 
The package \textsf{fdasrvf} \cite{tuckerFdasrvfElasticFunctional2023} implements methods for functional data registration. However, most packages primarily handle univariate functional data that are well-described by their coefficients in a given basis of functions. To address this limitation, the \textsf{funData} package \cite{happ-kurzObjectOrientedSoftwareFunctional2020} aims to provide a unified framework to handle univariate and multivariate functional data defined on different dimensional domains. Sparse functional data can also be considered in this package. Additionally, the \textsf{MFPCA} package \cite{happ-kurzMFPCAMultivariateFunctional2022}, built on top of the \textsf{funData} package, implements multivariate functional principal components analysis (MFPCA) for data defined on different dimensional domains \cite{happMultivariateFunctionalPrincipal2018}. For a complete overview of R packages related to FDA, refer to the CRAN webpage\footnote{\url{https://cran.r-project.org/web/views/FunctionalData.html}}.

In the Python community, there are relatively few packages specifically designed for FDA. Among these, \textsf{sktime} \cite{loningSktimeSktimeV02022} and \textsf{tslearn} \cite{tavenardTslearnMachineLearning2020} provide tools for the analysis of time series as a \textsf{scikit-learn} compatible API \cite{sklearn_api}. These packages implement specific time series methods such as Dynamic Time Warping (DTW) or shapelets learning, making them valuable for certain types of FDA. The primary Python package that focuses specifically on FDA is \textsf{scikit-fda} \cite{ramos-carrenoScikitfdaPythonPackage2023}. In particular, it implements diverse registration techniques as well as statistical data depths for functional data. However, most of the methods are designed for one-dimensional data and the package only supports multivariate functional data defined on the same domain. This gap presents an opportunity for the development of FDA tools for functional data defined on different dimensional domains and eventually sampled on irregular grids within the Python ecosystem.

This paper introduces \textsf{FDApy}: a package for the analysis of functional data in Python. The package aims to provide functionalities for creating and manipulating general functional data objects. \textsf{FDApy} thus supports the analysis of various types of functional data, whether densely or irregularly sampled, multivariate, or multidimensional. \textsf{FDApy} implements dimension reduction techniques \cite{ramsayFunctionalDataAnalysis2005,happMultivariateFunctionalPrincipal2018,golovkineClusteringMultivariateFunctional2022}, facilitating the extraction of  patterns from complex functional datasets. A large simulation toolbox, based on basis decomposition, is provided. It allows users to configure parameters for simulating different clusters within the data. Finally, some visualization tools are also available. Implementation of new methods, for regression or clustering for example, can easily be added.

\subsection{Data representation} 
\label{sub:data_representation}

In the general case, the data consist of independent trajectories of a vector-valued stochastic process $X = (X^{(1)}, \dots, X^{(P)})^\top$, $P\geq 1$. For each $1\leq p \leq P$, let $\TT_p \subset \RR^{d_p}$ with $d_p\geq 1$, as for instance, $\TT_p = [0,1]^{d_p}$. The realizations of each coordinate $X^{(p)}:\TT_p \rightarrow \RR$ are assumed to belong to  $\sLp{2}{\TT_p}$, the Hilbert space of squared-integrable real-valued functions defined on $\TT_p$. Thus $X$ is a stochastic process indexed by $\pointt = (t_1,\ldots,t_P)$ belonging to the $P-$fold Cartesian product $\TT \coloneqq \TT_1 \times \cdots\times \TT_P$ and taking values in the $P-$fold Cartesian product space $\HH \coloneqq \sLp{2}{\TT_1} \times \dots \times \sLp{2}{\TT_P}$. We denote by $\inH{\cdot, \cdot}$ the inner-product in $\HH$. In practice, realizations of functional data are only obtained on a finite grid, which is not necessarily the same for each realization and may include noise. Consider $N$ curves $X_1, \dots, X_n, \dots, X_N$ generated as a random sample of the $P$-dimensional stochastic process $X$ with continuous trajectories. For each $1 \leq n \leq N$, and given a vector of positive integers $\boldsymbol{M}_n = (M_n^{(1)}, \dots, M_n^{(P)}) \in \mathbb{R}^P$, let $T_{n, \boldsymbol{m}} = (T_{n, m_1}^{(1)}, \dots, T_{n, m_P}^{(P)}), 1 \leq m_p \leq M_n^{(p)}, 1 \leq p \leq P$, be the random observation times for the curve $X_n$. These times are obtained as independent copies of a variable $\boldsymbol{T}$ taking values in $\TT$. The vectors $\boldsymbol{M}_1, \dots, \boldsymbol{M}_N$ represent an independent sample of an integer-valued random vector $\boldsymbol{M}$ with expectation $\boldsymbol{\mu}_{\boldsymbol{M}}$. We assume that the realizations of $X$, $\boldsymbol{M}$ and $\boldsymbol{T}$ are mutually independent. The observations associated with a curve, or trajectory, $X_n$ consist of the pairs $(Y_{n, \boldsymbol{m}}, T_{n, \boldsymbol{m}}) \in \mathbb{R}^P \times \mathcal{T}$ where $\boldsymbol{m} = (m_1, \dots, m_P), 1 \leq m_p \leq M_n^{(p)}, 1 \leq p \leq P$, and $Y_{n, \boldsymbol{m}}$ is defined as
\begin{equation}\label{eq:model}
Y_{n, \boldsymbol{m}} = X_n(T_{n, \boldsymbol{m}}) + \varepsilon_{n, \boldsymbol{m}}, \quad , 1 \leq n \leq N,
\end{equation}
and $\varepsilon_{n, \boldsymbol{m}}$ are independent copies of a centered error random vector $\varepsilon \in \RR^P$ with finite variance. We use the notation $X_n(\boldsymbol{t})$ for the value at $\boldsymbol{t}$ of the realization $X_n$ of $X$. Univariate functional data refers to the case where $P = 1$. Multidimensional functional data refers to the case where $d_p > 1$ for some $p$.


\subsection{Motivating datasets} 
\label{sub:motivating_datasets}

In functional data analysis, the main challenge is to have a common representation for the different types of functional data that can be integrated into statistical models. We focus here on dimension reduction. We examine three functional datasets, each with distinct characteristics to address these challenges.

\begin{figure}
    \centering
    \subfloat[Temperature.]{\includegraphics[scale=0.6]{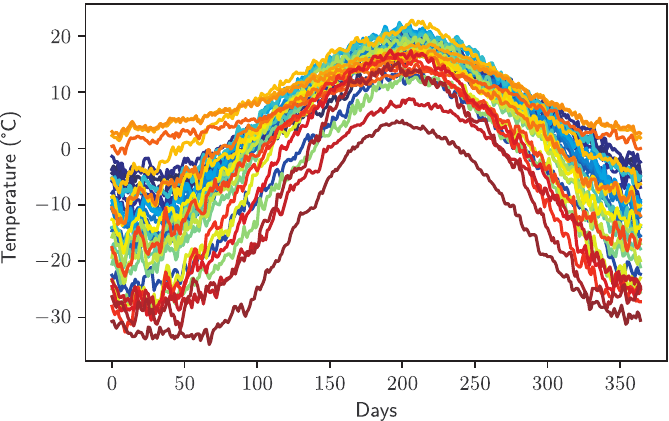}}
    \hfill
    \subfloat[Precipitation.]{\includegraphics[scale=0.6]{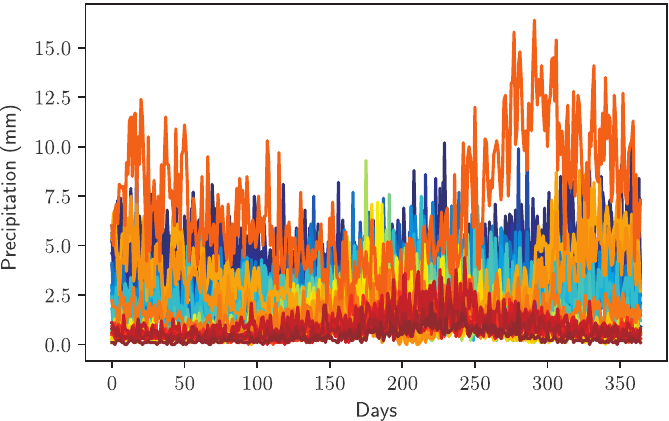}}
    \caption{Canadian weather dataset. Each color represents a different Canadian weather station.}
    \label{fig:canadian_weather}
\end{figure}

The Canadian weather dataset is presented in \cite{ramsayFunctionalDataAnalysis2005} and available in the R package \texttt{fda} \cite{ramsayFdaFunctionalData2023}. It contains daily recording of the temperature (in °C) and the precipitation (in mm, rounded to $0.1$mm) for $35$ Canadian cities. These recordings span each day of the year and are averaged over the years $1960$ to $1994$. This dataset is an example of multivariate functional (with $P = 2$) that includes measurement errors. Figure~\ref{fig:canadian_weather} presents an aggregated view of the data, showing temperature and precipitation patterns for various cities.

\begin{figure}
    \centering
    \subfloat[Albumin.]{\includegraphics[scale=0.4]{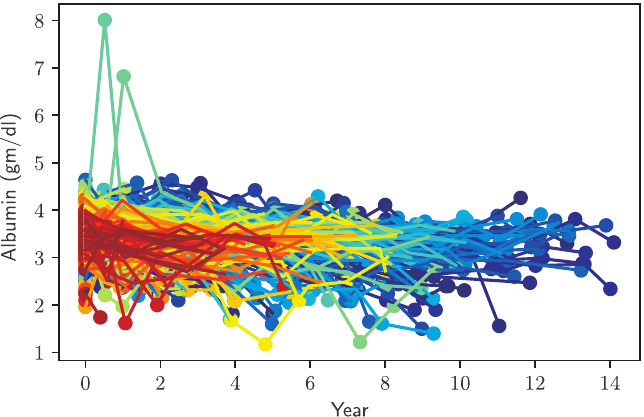}}
    \hfill
    \subfloat[Bilirubin.]{\includegraphics[scale=0.4]{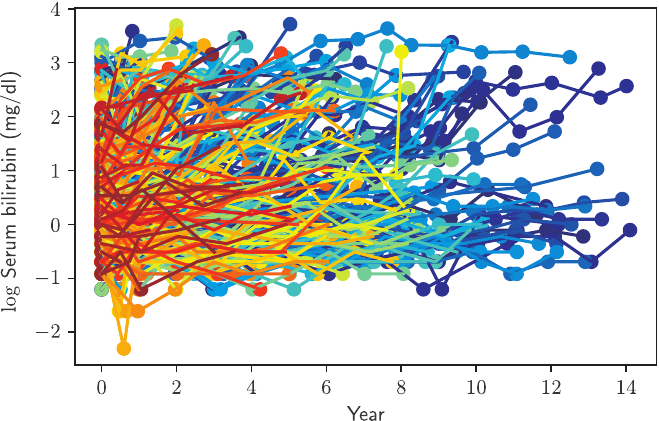}}
    \hfill
    \subfloat[Prothrombin.]{\includegraphics[scale=0.4]{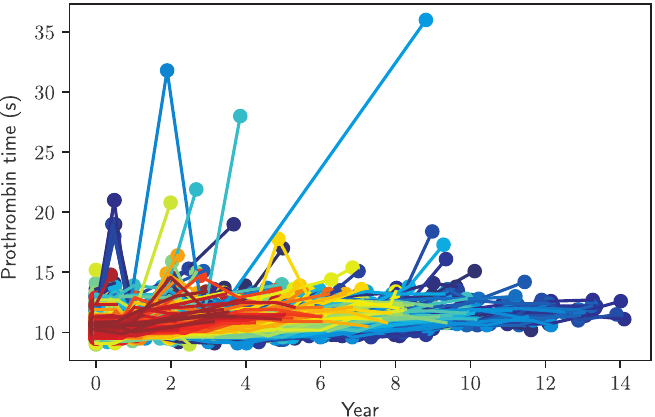}}
    \caption{Primary Biliary Cirrhosis dataset. Each color represents a different patient. Each point represents an observation for a given patient. The observation grid is different for each patient.}
    \label{fig:pbc}
\end{figure}

The Primary Biliary Cirrhosis (PBC) dataset \cite{murtaughPrimaryBiliaryCirrhosis1994} is a dataset from the Mayo Clinic containing measurements of various biomarkers from blood analysis of $312$ patients who have been diagnosed with primary cirrhosis of the liver, a rare autoimmune liver disease. We consider three biomarkers: albumin, bilirubin and prothrombin time, observed up to $15$ years after the first visit. We consider these biomarkers because they have been proven to be good predictors of patient outcomes \cite{sortFunctionalGeneralizedCanonical2023}. The dataset is available in the R package \texttt{JM} \cite{rizopoulosJMPackageJoint2010} and is an example of multivariate functional data (with $P =3$) with irregular sampling points. Figure~\ref{fig:pbc} presents an aggregated view of the data, showing the observation patterns for each biomarker across different patients.

\begin{figure}
    \centering
    \subfloat[Bogdanović.]{\includegraphics[scale=0.2]{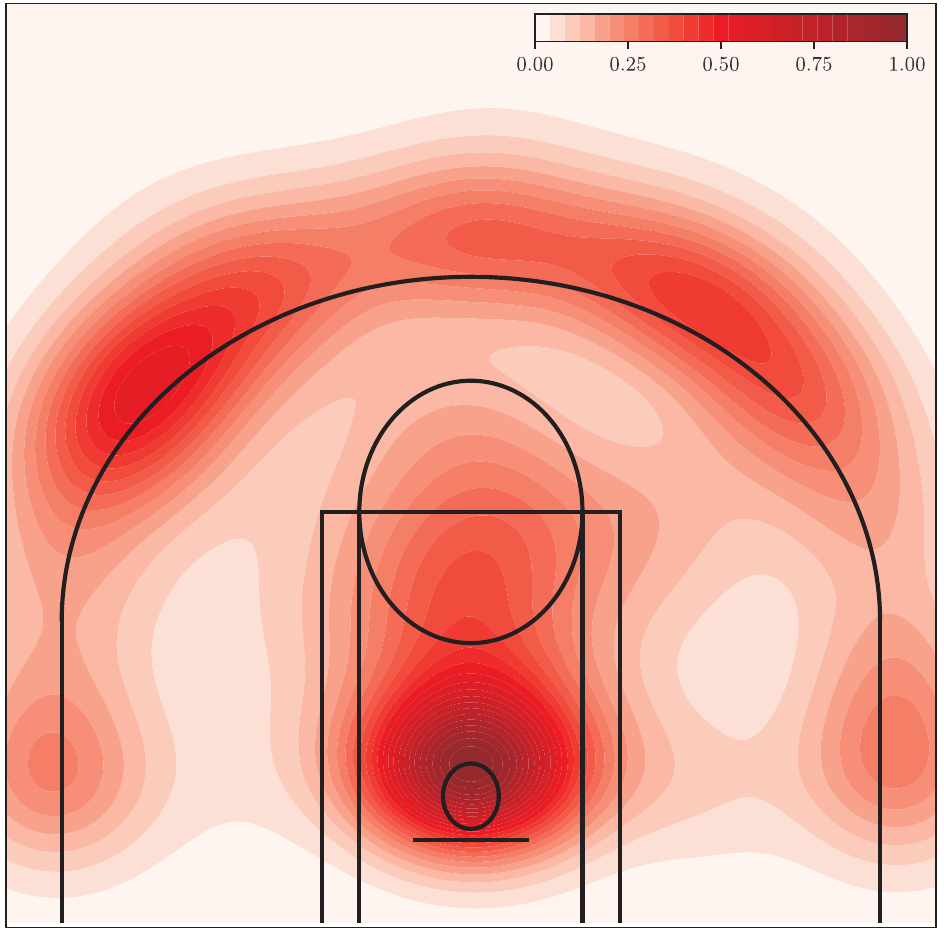}}
    \hfill
    \subfloat[Curry.]{\includegraphics[scale=0.2]{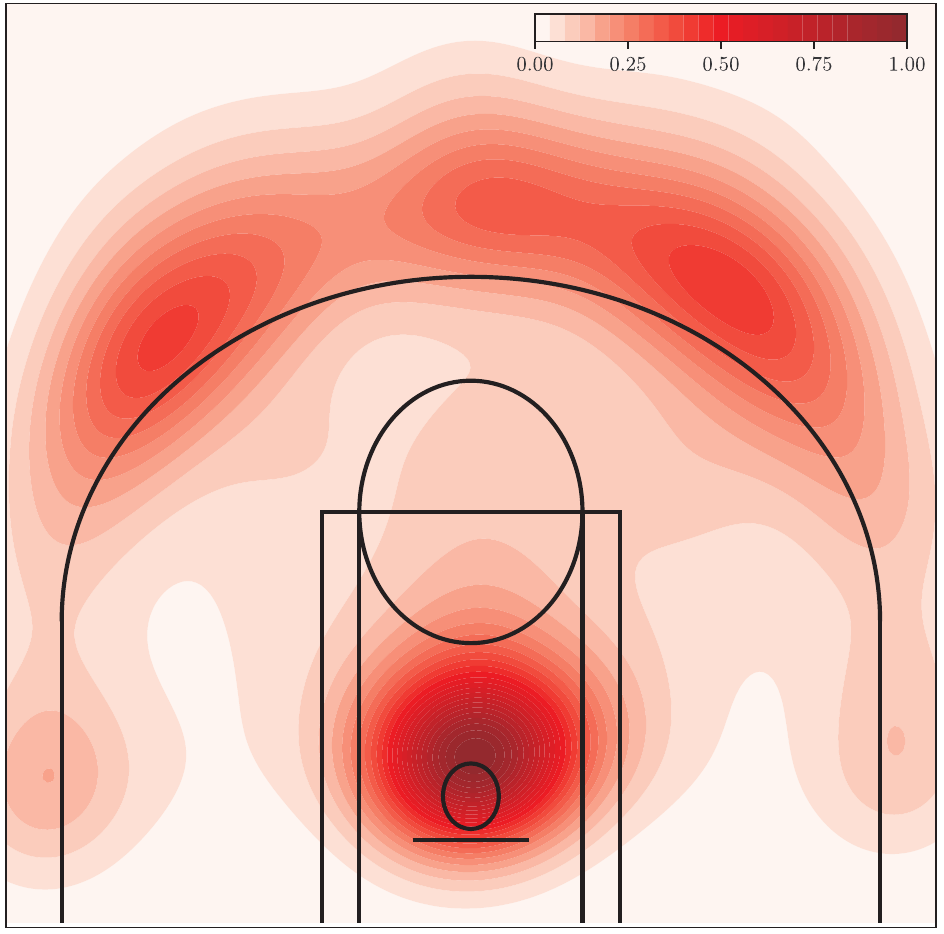}}
    \hfill
    \subfloat[James.]{\includegraphics[scale=0.2]{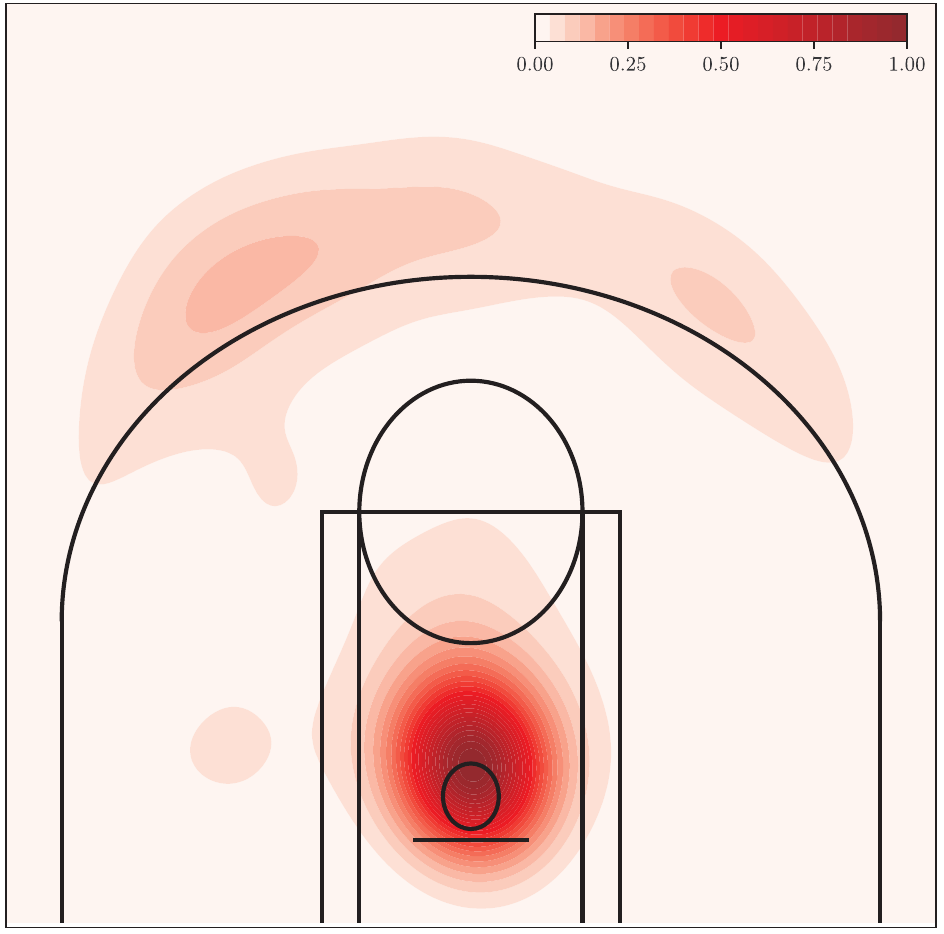}}
    \hfill
    \subfloat[Antetokounmpo.]{\includegraphics[scale=0.2]{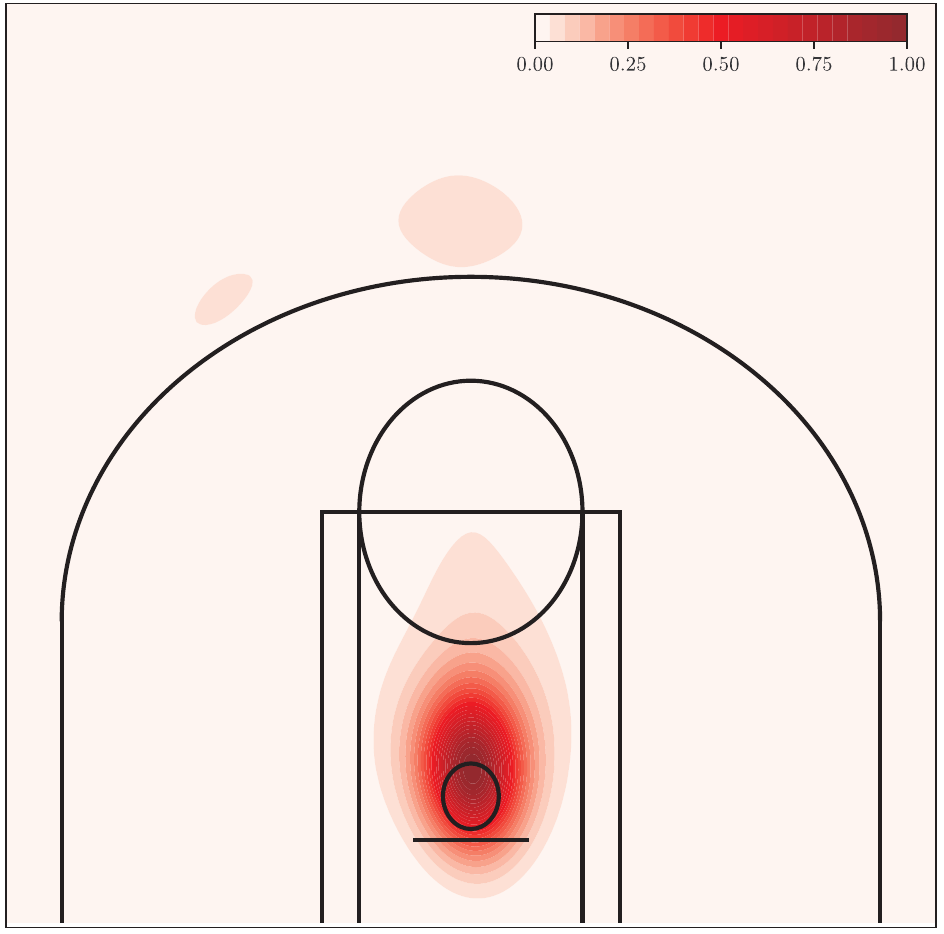}}
    \caption{NBA shooting dataset.}
    \label{fig:nba_shooting}
\end{figure}

The NBA shooting dataset contains density estimations of the shooting positions of $119$ players spanning the seasons between $2018 - 2019$ and $2022 - 2023$. The data have been retrieved from the APIs of \url{nba.com} using the Python package \texttt{nba\_api}\footnote{\url{https://github.com/swar/nba_api/}}. The data are presented more extensively in \cite{golovkineUseGramMatrix2023}. This dataset is an example of (univariate) multidimensional functional data ($d_1 = 2)$. Figure~\ref{fig:nba_shooting} presents the shooting patterns for four players with different behaviors (Bogdan Bogadanović, Stephen Curry, LeBron James and Giannis Antetokounmpo).

These datasets illustrate the variety and complexity of functional data, highlighting the challenges of their representation and analysis. The Canadian weather dataset showcases multivariate data with measurement errors, the PBC dataset presents multivariate data with irregular sampling, and the NBA shooting dataset provides an example of multidimensional data.



\section{Methodology for dimension reduction} 
\label{sec:methodology}

Dimension reduction techniques are crucial in functional data analysis due to the infinite-dimensional nature of functional data. Among these techniques, functional principal components analysis (FPCA) and multivariate functional principal components analysis (MFPCA) are particularly important. The goal of FPCA, for univariate functional data, and MFPCA, for multivariate functional data, is to represent the observations as a linear combination of basis functions, which capture the directions of maximum variation within the data.

Let $\mu : \TT{} \rightarrow \HH$ denote the mean function of the process $X$, defined as $\mu(\pointt) \coloneqq \EE(X(\pointt)),\,\pointt \in \TT{}$. The covariance function $C$ is the $P \times P$ matrix-valued function defined, for $\points, \pointt \in \TT{}$, as
\begin{equation}\label{eq:covariance_function}
    C(\points, \pointt) \coloneqq \EE\left(\{X(\points) - \mu(\points)\}\{X(\pointt) - \mu(\pointt)\}^{\top}\right), \quad \points, \pointt \in \TT{}.
\end{equation}
To be more specific, for $1 \leq p, q \leq P$, the $(p, q)$th entry of the matrix $C(\points, \pointt)$ is the covariance function between the $p$th and the $q$th features of the process $X$:
\begin{equation}\label{eq:covariance_function_components}
    C_{p, q}(s_p, t_q) \coloneqq \EE\left(\{\Xp{p}(s_p) - \mup{p}(s_p)\}\{\Xp{q}(t_q) - \mup{q}(t_q)\}\right), \quad s_p \in \TT_{p}, t_q \in \TT_{q}.
\end{equation}
The covariance operator $\Gamma : \HH \rightarrow \HH$ of $X$, defined as an integral operator with kernel $C$, is a compact positive operator on $\HH$, therefore it is diagonalizable. Using the multivariate Karhunen-Loève theorem \cite{happMultivariateFunctionalPrincipal2018}, the process $X$ can be decomposed as
\begin{equation}\label{eq:kl_multi}
    X(\pointt) = \mu(\pointt) + \sum_{k = 1}^\infty \mathfrak{c}_k \phi_k(\pointt), \quad \pointt \in \TT{},
\end{equation}
where $\phi_k$ are the eigenfunctions and $\mathfrak{c}_{k} = \inH{X - \mu, \phi_k}$ are the projections of the centered curves onto the eigenfunctions. In practice, this expansion is truncated to a finite number of terms as the observations are sampled on a finite grid. Let
\begin{equation}\label{eq:kl_multi_trunc}
    X_{\lceil K \rceil}(\pointt) = \mu(\pointt) + \sum_{k = 1}^K \mathfrak{c}_k \phi_k(\pointt), \quad \pointt \in \TT{}, \quad K \geq 1,
\end{equation}
be the truncated Karhunen-Loève expansion of the process $X$ and
\begin{equation}\label{eq:kl_multi_trunc_comp}
    X_{\lceil K_p \rceil}^{(p)}(t_p) = \mup{p}(t_p) + \sum_{k = 1}^{K_p} \xi^{(p)}_k \varphi_k^{(p)}(t_p), \quad t_p \in \TT{p}, \quad K_p \geq 1, \quad 1 \leq p \leq P,
\end{equation}
be the truncated Karhunen-Loève expansion of the $p$th feature of the process $X$. For each $p$, the function $\mup{p}$ is the $p$th feature of the multivariate mean function $\mu$ and the set $\{\varphi^{(p)}_k\}_{1 \leq k \leq K_p}$ is a basis of univariate functions in $\sLp{}{\TT{p}}$, whose elements are not the components of the multivariate functions $\phi_k$. In~\eqref{eq:kl_multi_trunc_comp}, the coefficients $\xi^{(p)}_k$ are the projection of the centered curve $\Xp{p}$ onto the eigenfunctions $\varphi_k^{(p)}$ and are not (directly) related to the coefficients $\mathfrak{c}_k$ in~\eqref{eq:kl_multi_trunc}.

We present MFPCA for functional data that may be multivariate, multidimensional and eventually irregularly sampled. We illustrate two methods to estimate the multivariate eigencomponents: diagonalization of the covariance operator and diagonalizaion of the inner-product matrix of the multivariate functional data. These two methods are theoretically equivalent but some differences may appear in practice (see \cite{golovkineUseGramMatrix2023} for details). 

\subsection{Diagonalization of the covariance operator} 
\label{sub:by_diagonalization_of_the_covariance_operator}

The estimation of the eigencomponents of the covariance $\Gamma$ by its diagonalization is derived in \cite{happMultivariateFunctionalPrincipal2018} for a general class of multivariate functional data defined on different dimensional domains. his method establishes a direct relationship between the univariate truncated representation \eqref{eq:kl_multi_trunc_comp} of the univariate elements $X^{(p)}$ and the multivariate truncated representation \eqref{eq:kl_multi_trunc} of the multivariate functional data $X$.

We recall here how to estimate the eigencomponents. Following \cite[Prop.~5]{happMultivariateFunctionalPrincipal2018}, the multivariate components for $X$ are estimated by a weighted combination of the univariate components computed from each $X^{(p)}$. First, we perform an univariate FPCA on each of the features of $X$ separately (see, e.g., \cite{ramsayFunctionalDataAnalysis2005,yaoFunctionalDataAnalysis2005,allenMultiwayFunctionalPrincipal2013}). For each $p$, this results in a set of eigenfunctions $\{\varphi_k^{(p)}\}_{1 \leq k \leq K_p}$ associated with a set of eigenvalues $\{\lambda_k^{(p)}\}_{1 \leq k \leq K_p}$ for a given truncation integer $K_p$. The univariate scores for a realization $\Xnp$ of $X^{(p)}$ are given by $\xi_{nk}^{(p)} = \inLp{\Xnp, \varphi_k^{(p)}}, ~1 \leq k \leq K_p$. Considering $K_+ \coloneqq \sum_{p = 1}^P K_p$, we then define the matrix $\mathcal{Z} \in \mathbb{R}^{N \times K_+}$, the concatenation of the scores. An estimation of the covariance of the matrix $\mathcal{Z}$ is given by $\mathbf{Z} = (N - 1)^{-1}\mathcal{Z}^\top\mathcal{Z}$. An eigenanalysis of the matrix $\mathbf{Z}$ is carried out to estimate the eigenvectors $\boldsymbol{v}_k$ and eigenvalues $\lambda_k$. Finally, the multivariate eigenfunctions are estimated as a linear combination of the univariate eigenfunctions using
\begin{equation*}
\phi_k^{(p)}(t_p) = \sum_{l = 1}^{K_p}[\boldsymbol{v}_k]_{l}^{(p)}\varphi_{l}^{(p)}(t_p),\quad t_p \in \TT_{p},\quad 1 \leq k \leq K_+,\quad 1 \leq p \leq P,
\end{equation*}
where $[\boldsymbol{v}_k]^{(p)}_{l}$ denotes the $l$th entry of the $p$th block of the vector $\boldsymbol{v}_k$. The multivariate scores are estimated as
$$\mathfrak{c}_{nk} = \mathcal{Z}_{{n,\cdot}}\boldsymbol{v}_k, \quad 1 \leq n \leq N, \quad 1 \leq k \leq K_+,$$
where $\mathcal{Z}_{{n,\cdot}}$ is the $n$th row of the matrix $\mathcal{Z}$.


\subsection{Diagonalization of the inner product matrix} 
\label{sub:by_diagonalization_of_the_inner_product_matrix}

The estimation of the eigencomponents of the covariance $\Gamma$ using the inner-product matrix of the functional dataset is explained in \cite{golovkineUseGramMatrix2023}. 
Consider the inner-product matrix $\mathbf{M}$, with entries defined as
\begin{equation}\label{eq:gram_mat}
    \mathbf{M}_{nn^\prime} = \frac{1}{N}\inH{X_n - \mu, X_{n^\prime} - \mu}, \quad n, n^\prime = 1, \dots, N.
\end{equation}
Let $\{l_k\}_{1 \leq k \leq N}$ such that $l_1 \geq \dots \geq l_N \geq 0$ be the set of eigenvalues and $\{\boldsymbol{u}_k\}_{1 \leq k \leq N}$ be the set of eigenvectors of the matrix $\mathbf{M}$. The relationship between all nonzero eigenvalues of the covariance operator $\Gamma$ and the eigenvalues of $\mathbf{M}$ is given by
\begin{equation}\label{eq:eigenvalues_relation_p}
    \lambda_k = l_k, \quad k = 1, 2, \dots, N,
\end{equation}
while the relationship between the multivariate eigenfunctions of the covariance operator $\Gamma$ and the orthonormal eigenvectors of $\mathbf{M}$ is given by
\begin{equation}\label{eq:eigenfunction_relation_p}
    \phi_k(\pointt) = \frac{1}{\sqrt{N l_k}}\sum_{n = 1}^N [\boldsymbol{u}_{k}]_n\left\{X_n(\pointt) - \mu(\pointt)\right\}, \quad \pointt \in \TT{}, \quad k = 1, 2, \dots, N, 
\end{equation}
where $[\boldsymbol{u}_{k}]_n$ is the $n$th entry of the vector $\boldsymbol{u}_k$. The scores are then computed as the inner-product between the multivariate curves and the multivariate eigenfunctions and are given by
\begin{equation}\label{eq:scores_relation_p}
    \mathfrak{c}_{nk} = \sqrt{N l_k}[\boldsymbol{u}_{k}]_n, \quad n = 1, 2, \dots, N, \quad k = 1, 2, \dots, N. 
\end{equation}



\section{Implementation} 
\label{sec:implementation}

Implementing functional data analysis techniques requires a robust and flexible software framework. We describe the object-oriented approach we use in Python to represent functional data objects and perform MFPCA.

\subsection{Representation of functional data} 
\label{sub:representation_functional_data}

We use an object-oriented approach to represent functional data in Python, defining functional data as objects. Objects consists of two main components: data and methods. The data can be of various types, such as lists or matrices, while the methods are functions designed to access, modify, and interact with the data and other objects. These objects can then be seamlessly integrated into the implementation of statistical models for functional data, similar to how matrices are used in statistical models for multivariate data.

In Python, objects are created using classes. The core of the package is the abstract class \texttt{FunctionalData}. Two classes extends this abstract class: \texttt{BasisFunctionalData} and \texttt{GridFunctionalData}. The class \texttt{BasisFunctionalData} represents functional data using basis expansions, while the class \texttt{GridFunctionalData} represents functional data using sampling points. Two subclasses further extend \texttt{GridFunctionalData}:
\begin{enumerate}
    \item Class \texttt{DenseFunctionalData} represents dense functional data of arbitrary dimension (one for curves, two for images, etc.) on a common set of sampling points $t_1, \dotsc, t_M$ for all the observations.
    \item Class \texttt{IrregularFunctionalData} represents irregular functional data of arbitrary dimension on different sets of sampling points. The number and the location of the sampling points vary between observations.
\end{enumerate}
A schematic representation of the class hierarchy is given in Figure \ref{fig:schema}.

\begin{figure}
\centering
\begin{tikzpicture}[
    nonterminal/.style={
      rectangle, minimum size=6mm, very thick,
      draw=red!50!black!50,
      font=\ttfamily
    }]
    \matrix[row sep=1mm,column sep=5mm] {
      & \node(bfd)[nonterminal]{BasisFunctionalData}; &  \\
    \node(fd)[nonterminal] {FunctionalData}; & & \node(dfd)[nonterminal]{DenseFunctionalData}; \\
      & \node(gfd)[nonterminal]{GridFunctionalData}; &  \\
      & & \node(ifd)[nonterminal]{IrregularFunctionalData}; \\
  };
  
  \draw[->,>=latex] (fd.north) |- (bfd.west);
  \draw[->,>=latex] (fd.south) |- (gfd.west);
  \draw[->,>=latex] (gfd.north) |- (dfd.west);
  \draw[->,>=latex] (gfd.south) |- (ifd.west);
\end{tikzpicture}
\caption{Schematic representation of the class hierarchy.}
\label{fig:schema}
\end{figure}

The \texttt{MultivariateFunctionalData} class represents multivariate functional data as a list of \texttt{FunctionalData} objects. This allows for the combination of different types of functional data (\texttt{DenseFunctionalData}, \texttt{IrregularFunctionalData}, \texttt{BasisFunctionalData}) within a single multivariate dataset. We can also mix unidimensional data (curves) with multidimensional data (images). This class has access to all the methods applicable to lists such as \texttt{append}, \texttt{extend}, \texttt{pop}, etc.

\begin{remark}\label{rem:coercion}
In practice, the difference between dense and irregularly sampled functional data can be subtle. By design, dense functional data are assumed to be sampled on the complete grid $\mathcal{T} = \{t_1, \dotsc, t_M\}$ and may include measurement errors. For example, sensors data recorded at a given sampling rate may have anomalies during the recording process. In contrast, irregularly sampled functional data are observed at different sampling points with varying number of points. This is usually the case in medical studies such as growth curves analysis because one cannot expect that the individuals are measured at the exact same time.
\end{remark}

The classes \texttt{DenseFunctionalData} and \texttt{IrregularFunctionalData} represent the data in a similar way: the instance variable \texttt{argvals} contains the sampling points and the instance variable \texttt{values} represents the data. In the case of dense functional data, the \texttt{argvals} argument is a dictionary, where each entry contains a \textbf{numpy} array that represents the common sampling points for a given dimension, while the \texttt{values} is a \textbf{numpy} array containing the observations. In the case of one dimensional data sampled on a grid with $M$ points, \texttt{argvals} contains only one entry as an array of shape $(M,)$ and \texttt{values} is an array of dimension $(N, M)$ where each row is an observation. For two dimensional observations with $M^{(1)} \times M^{(2)}$ sampling points, \texttt{argvals} contains two entries, the first is an array of shape $(M^{(1)},)$ and the second an array of shape $(M^{(2)},)$, and \texttt{values} is an array of dimension $(N, M^{(1)}, M^{(2)})$ where the first coordinate gives the observations. Higher dimensional data are represented by adding an entry in the \texttt{argvals} dictionary and a dimension in the \texttt{values} array. For irregularly sampled functional data, both \texttt{argvals} and \texttt{values} are dictionaries. The entries of \texttt{argvals} are dictionaries where each entry is the sampling points for a particular observation. In a similar way, each entry of the \texttt{values} dictionary represents an observation. For one dimensional irregularly sampled functional data, \texttt{argvals} contains one entry which is a dictionary of size $N$ containing the sampling points as array of shape $(M_n,), 1 \leq n \leq N$ and \texttt{values} is a dictionary with $N$ entries containing the observations as arrays of shape $(M_n,), 1 \leq n \leq N$. For higher dimensions, each entry of the \texttt{argvals} dictionary represents a dimension of the process and contains another dictionary with $N$ entries for the sampling points. Likewise, the \texttt{values} dictionary has $N$ entries and every one of them is an array of shape $(M^{(1)}_n, M^{(2)}_n, \dotsc), 1 \leq n \leq N$.


\subsection{Multivariate functional principal component analysis} 
\label{sub:multivariate_functional_principal_component_analysis}

The implementation of the MFPCA algorithm is encapsulated in the \texttt{MFPCA} class. This class requires an object of class \texttt{MultivariateFunctionalData} to perform the analysis, along with parameters specifying the number of principal components to retain (or the percentage of variance explained) and the method for computing the principal components. The available methods are
\begin{itemize}
    \item \texttt{covariance}: decomposition of the covariance operator, as detailed in Section~\ref{sub:by_diagonalization_of_the_covariance_operator} and \cite{happMultivariateFunctionalPrincipal2018};
    \item \texttt{inner-product}: decomposition of the inner-product matrix, as detailed in  Section~\ref{sub:by_diagonalization_of_the_inner_product_matrix} and \cite{golovkineUseGramMatrix2023}.
\end{itemize}
For the covariance decomposition method, an additional parameter specifies the approach for each univariate representation, which can be obtained via univariate FPCA or basis expansions (e.g., P-splines smoothing \cite{eilersPracticalSmoothingJoys2021}).



\section{Application} 
\label{sec:application}

We apply MFPCA to the three datasets presented in Section~\ref{sub:motivating_datasets}: Canadian weather, Primary Biliary Cirrhosis, and NBA shooting datasets. These applications illustrate the effectiveness of our software in analyzing complex, high-dimensional functional data.

\subsection{Canadian weather} 
\label{sub:canadian_weather}

\begin{figure}
    \centering
    \subfloat[Mean.]{
        \includegraphics[scale=0.36]{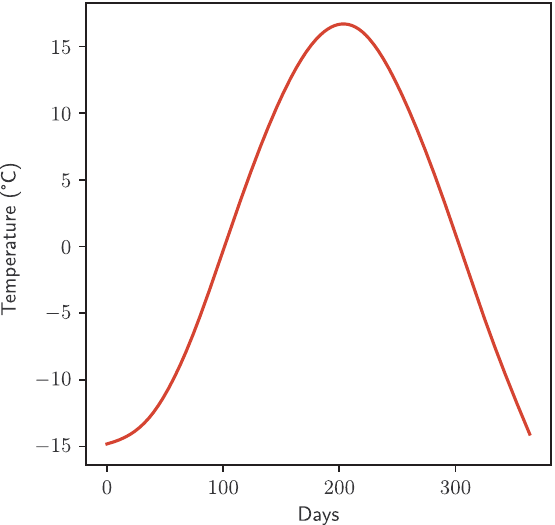}
        \includegraphics[scale=0.36]{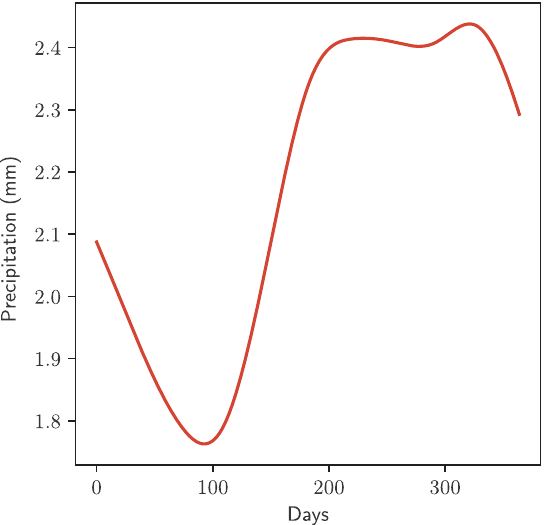}
    }
    \hfill
    \subfloat[Covariance.]{
        \includegraphics[scale=0.36]{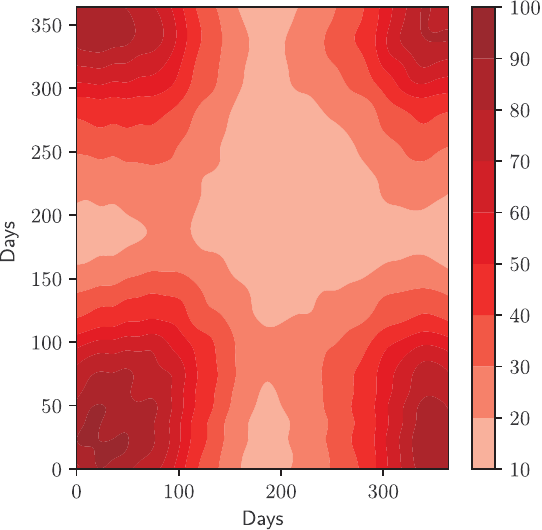}
        \includegraphics[scale=0.36]{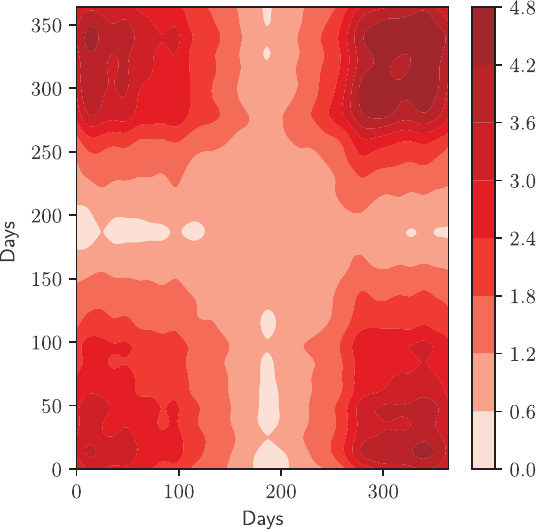}
    }
    \caption{Canadian weather dataset mean and covariance.}
    \label{fig:canadian_weather_mean_cov}
\end{figure}

We aim to estimate the mean and covariance functions, as well as the eigenfunctions of the Canadian weather dataset. The mean and covariance functions are estimated using P-splines smoothing \cite{eilersPracticalSmoothingJoys2021} with a penalty set to $1$. The penalty can be specified in the software. Figure~\ref{fig:canadian_weather_mean_cov} presents the estimation of the mean and the covariance for the Canadian weather dataset. The mean temperature reaches a peak around the middle of the year (end of June), while the mean precipitation is roughly constant for the second half of the year. Concerning the covariance, both the temperature and the precipitation exhibit similar patterns, with most of the variation occurring at the beginning and the end of the year.

\begin{figure}
    \centering
    \subfloat[Temperature.]{
        \includegraphics[scale=0.6]{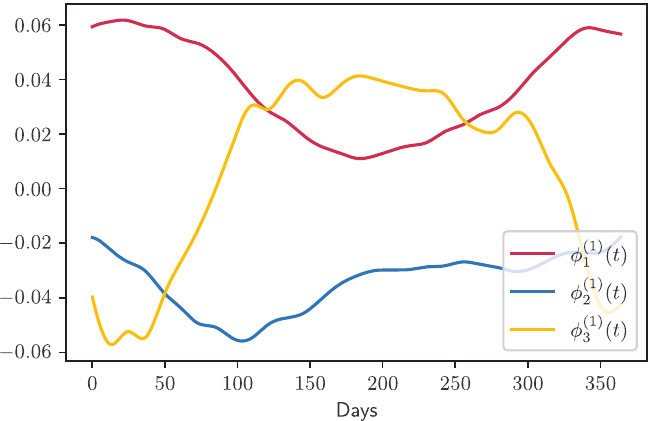}
    }
    \hfill
    \subfloat[Precipitation.]{
        \includegraphics[scale=0.6]{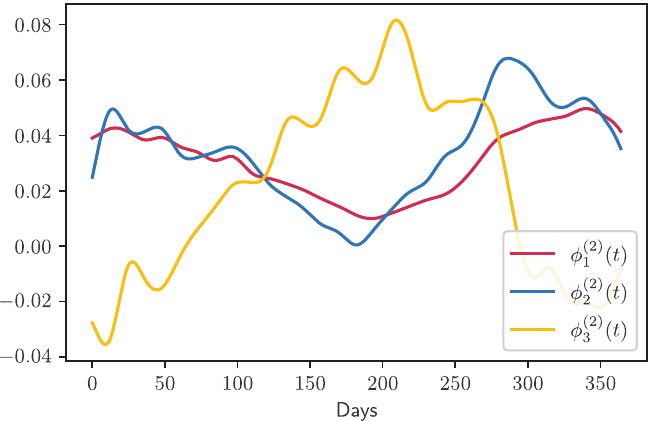}
    }
    \caption{Canadian weather dataset eigenfunctions.}
    \label{fig:canadian_weather_eigenfunctions}
\end{figure}

Figure~\ref{fig:canadian_weather_eigenfunctions} presents the results of the MFPCA, retaining the associated number of principal components such that $95\%$ of the variance is explained. The analysis was performed using univariate FPCA with $15$ principal components for each component. The number of univariate principal components is the choice of the user and can be specified in the software. The first principal component is positive for both features, indicating that weather stations with positive scores will have higher temperatures and more precipitation than average. This effect is more pronounced in winter than in summer because the eigenfunctions are closer to zero during summer. The other eigenfunctions can be interpreted similarly. 


\subsection{Primary Biliary Cirrhosis} 
\label{sub:primary_biliary_cirrhosis}

\begin{figure}
    \centering
    \subfloat[Mean]{
        \includegraphics[scale=0.4]{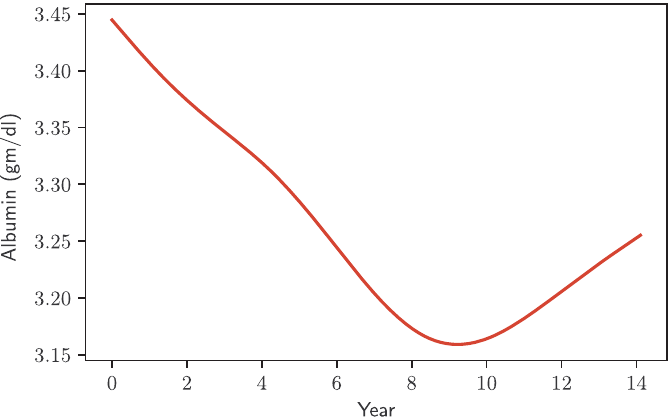}
        \includegraphics[scale=0.4]{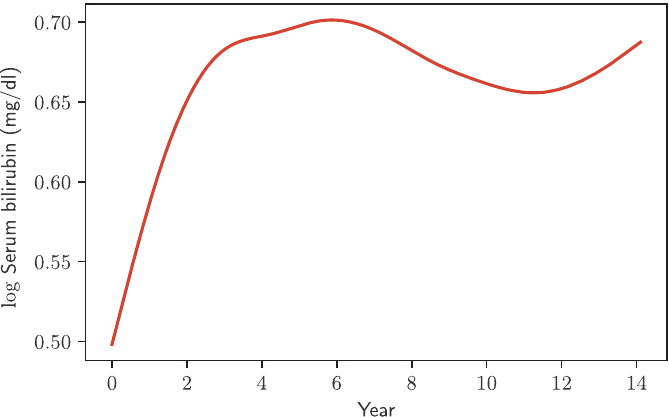}
        \includegraphics[scale=0.4]{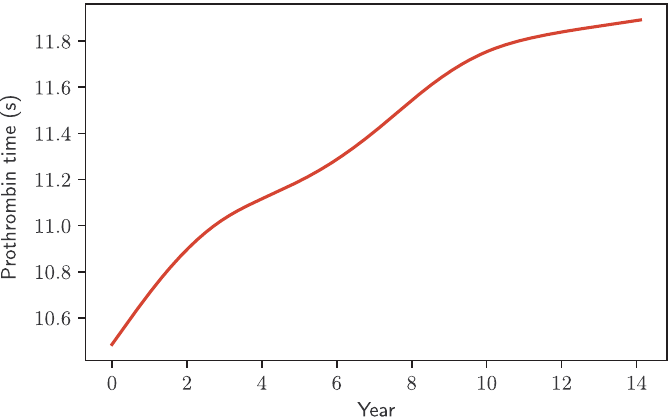}
    }
    \\
    \subfloat[Covariance]{
        \includegraphics[scale=0.5]{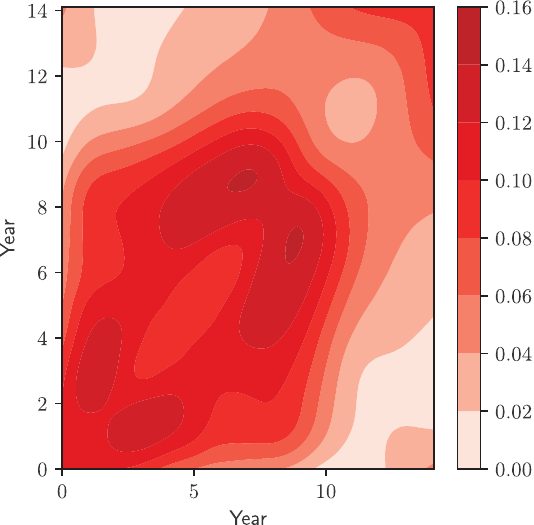}
        \includegraphics[scale=0.5]{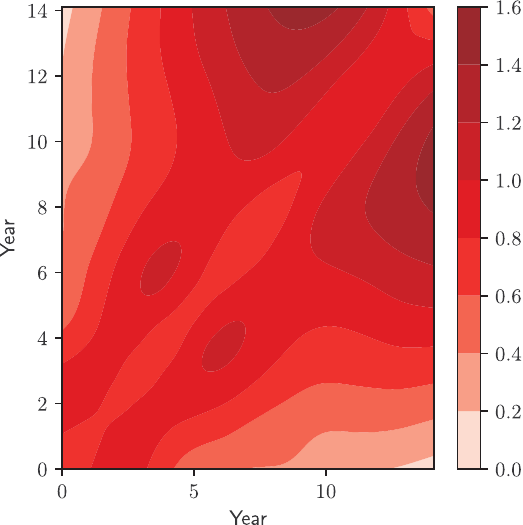}
        \includegraphics[scale=0.5]{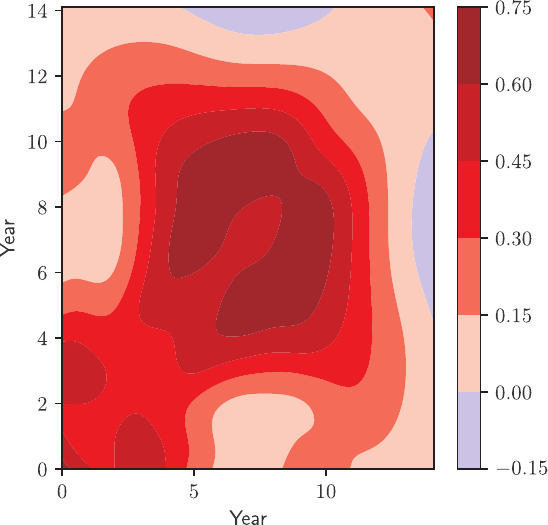}
    }
    \caption{Primary Biliary Cirrhosis dataset mean and covariance.}
    \label{fig:pbc_mean_cov}
\end{figure}

We aim to estimate the mean and covariance functions, as well as the eigenfunctions of the Primary Biliary Cirrhosis dataset. The mean and covariance functions are estimated using P-splines smoothing \cite{eilersPracticalSmoothingJoys2021} with a large penalty set to $100$ due to the scarcity of data. Figure~\ref{fig:pbc_mean_cov} presents the mean and covariance functions for the dataset. The mean albumin decreases until $10$ years after the first visit and increases afterward. The mean bilirubin and mean prothrombin increase over time. 

\begin{figure}
    \centering
    \subfloat[Albumin]{
        \includegraphics[scale=0.4]{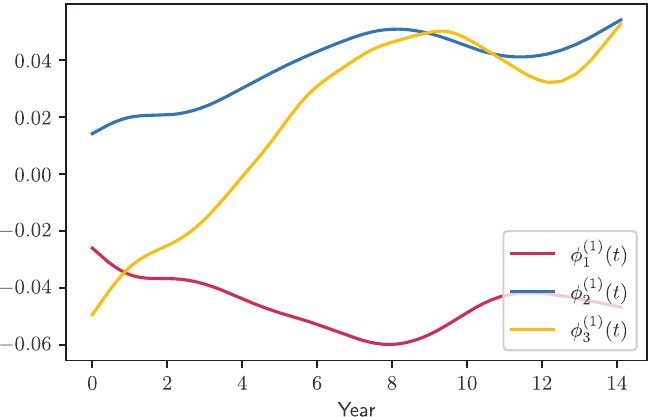}
    }
    \hfill
    \subfloat[Bilirubin]{
        \includegraphics[scale=0.4]{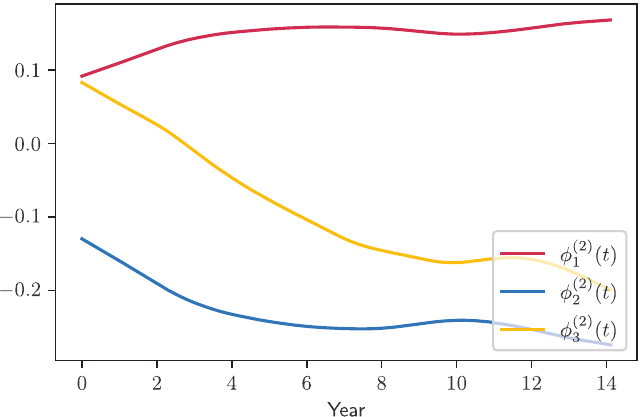}
    }
    \hfill
    \subfloat[Prothrombin]{
        \includegraphics[scale=0.4]{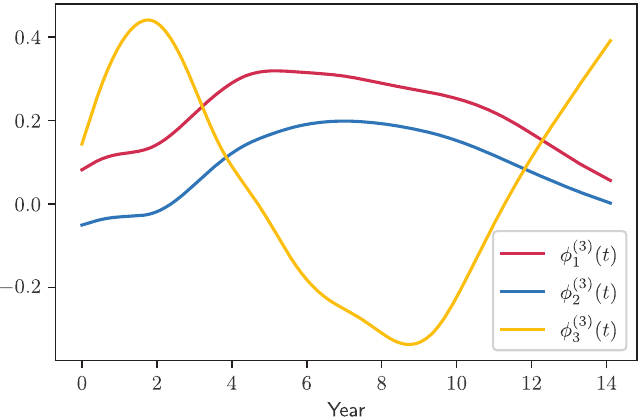}
    }
    \caption{Primary Biliary Cirrhosis dataset eigenfunctions.}
    \label{fig:pbc_eigenfunctions}
\end{figure}

Figure~\ref{fig:pbc_eigenfunctions} presents the results of the MFPCA with three principal components. The analysis was performed using univariate FPCA with five principal components for each feature. The first two principal components are approximately constant for all features, indicating that the differences in the biomarkers' trajectories between subjects are mostly due to an overall shift around the mean. The third component shows an increasing trend for albumin, a decreasing trend for bilirubin and a U-shaped for prothrombin. 


\subsection{NBA shoots} 
\label{sub:nba_shoots}

\begin{figure}
    \centering
    \subfloat[$\phi_1(t)$]{\includegraphics[scale=0.2]{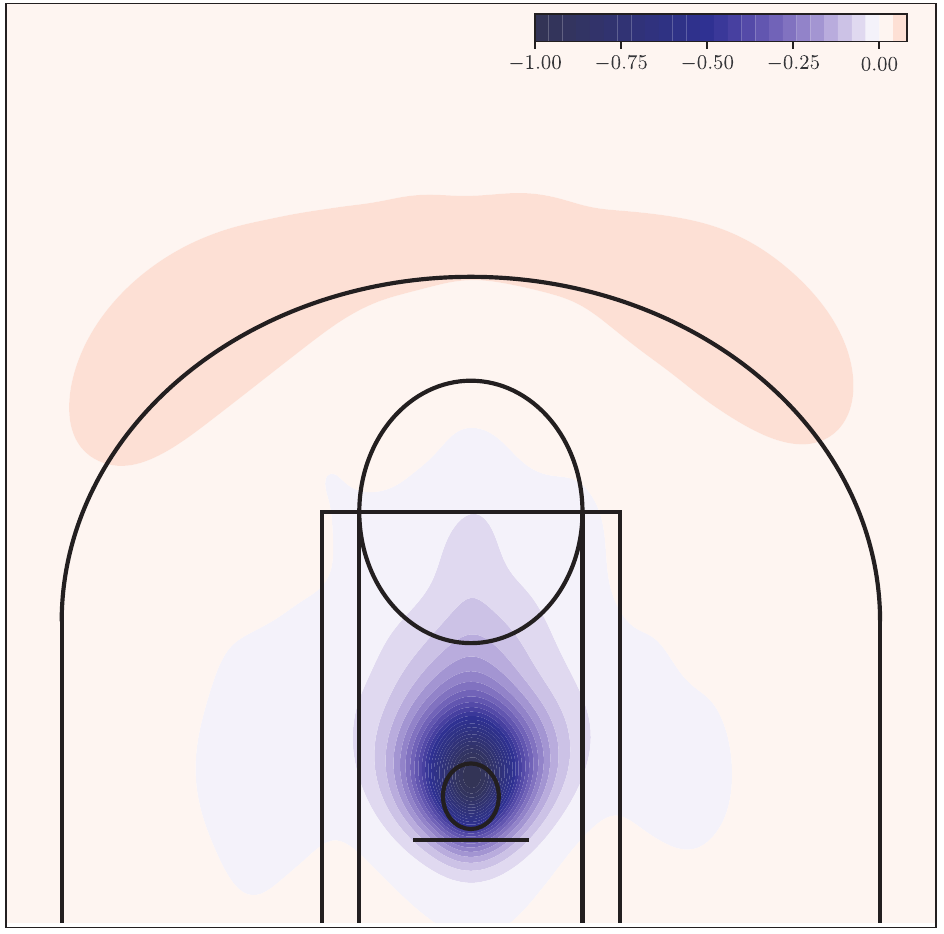}}
    \hfill
    \subfloat[$\phi_2(t)$]{\includegraphics[scale=0.2]{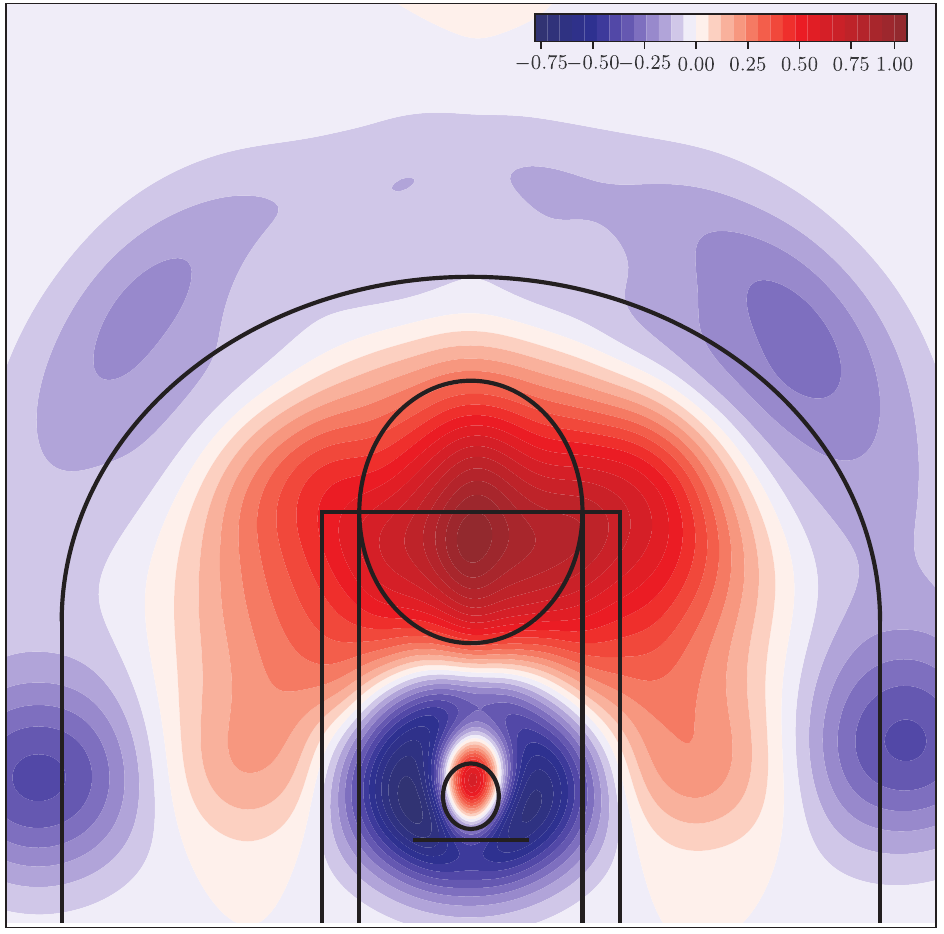}}
    \hfill
    \subfloat[$\phi_3(t)$]{\includegraphics[scale=0.2]{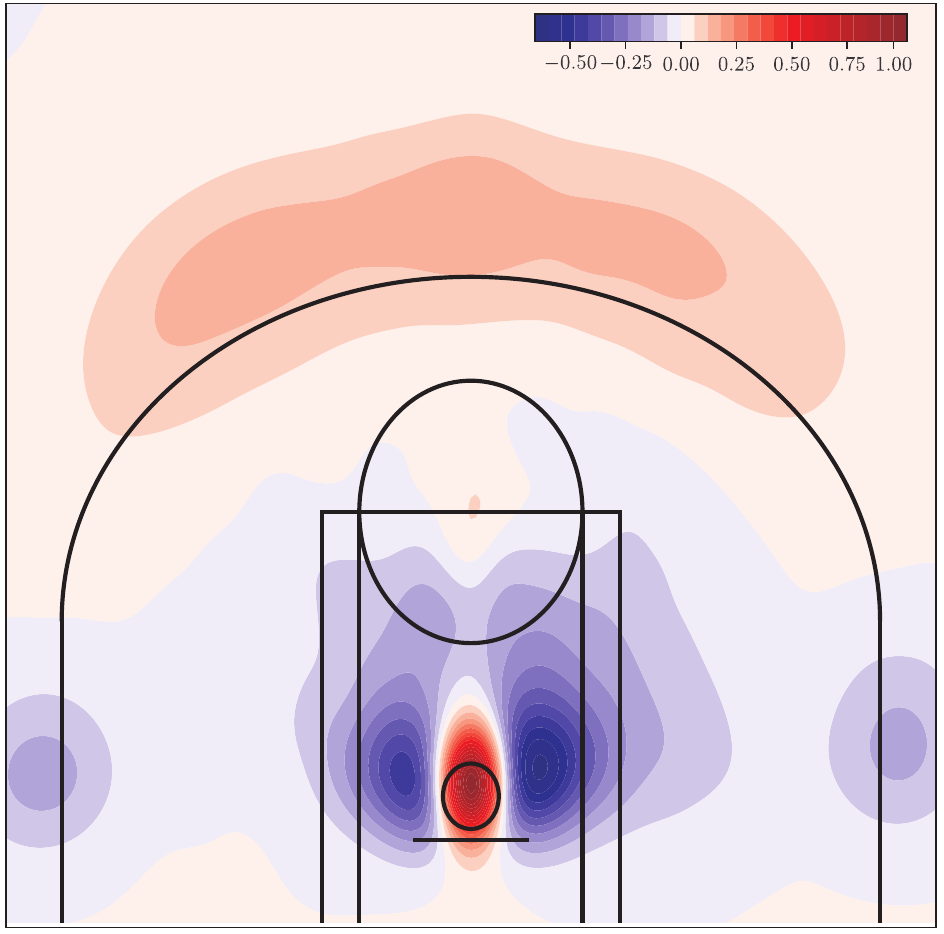}}
    \hfill
    \subfloat[$\phi_4(t)$]{\includegraphics[scale=0.2]{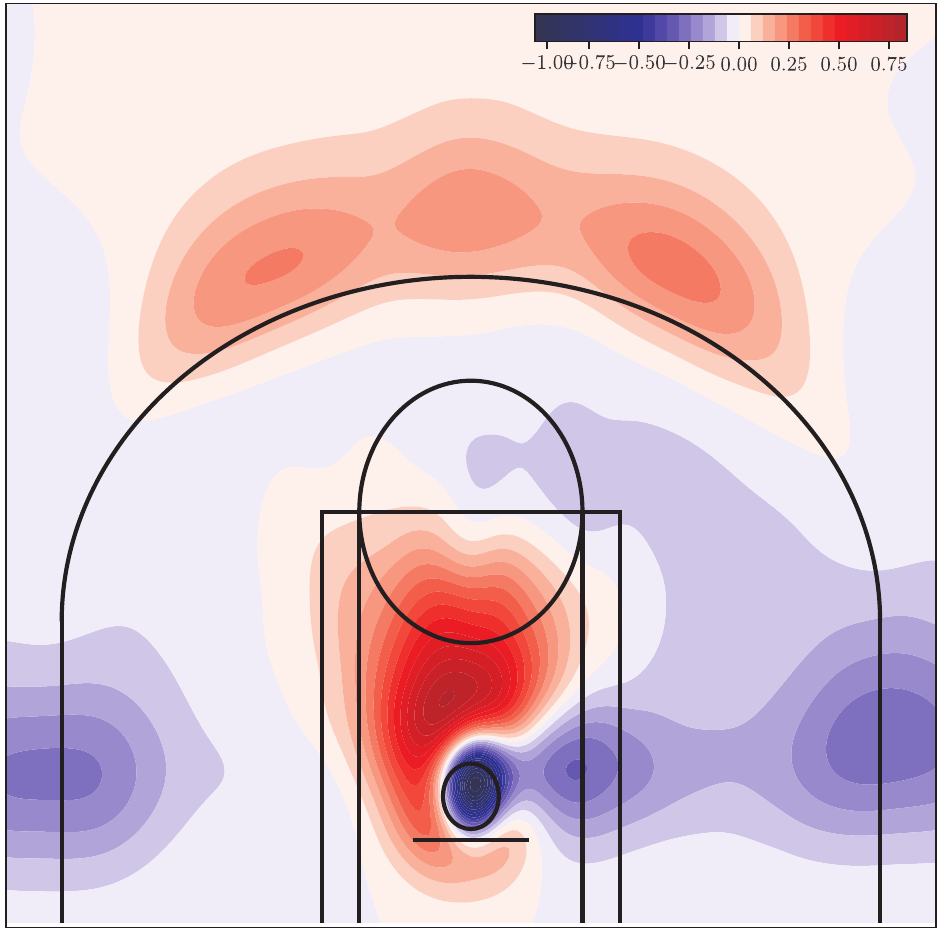}}
    \caption{NBA shooting eigenfunctions.}
    \label{fig:nba_shooting_eigenfunctions}
\end{figure}

We aim to analyze the NBA shooting dataset using MFPCA to understand sources of variation in shooting patterns among NBA players. Figure~\ref{fig:nba_shooting_eigenfunctions} presents the results of the MFPCA with four principal components. The analysis was performed using the inner-product matrix of the data. The first source of variation comes from shooting under the basket. Players who shoot frequently under the basket will tend to have large negative scores (recall that the eigenfunctions are defined up to a change of sign). The second eigenfunction differentiates players who shoot within and outside the three-points line. The third component also concerns the area under the basket. Finally, the fourth eigecomponent differentiates between shooting in the axis of the basket and the corners.



\section{Simulation} 
\label{sec:simulation}

\begin{figure}
    \centering
    \subfloat[First feature - Images.]{
        \includegraphics[scale=0.39]{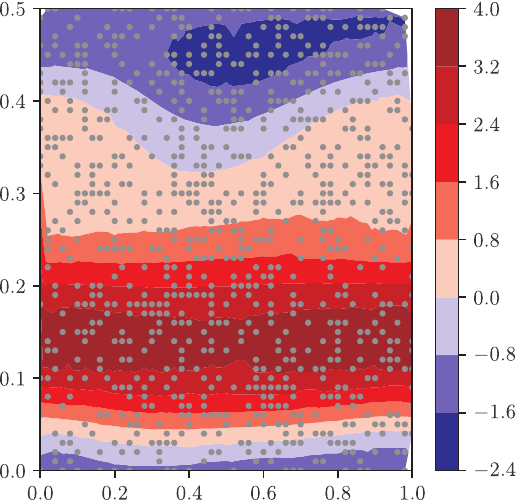}
        \includegraphics[scale=0.39]{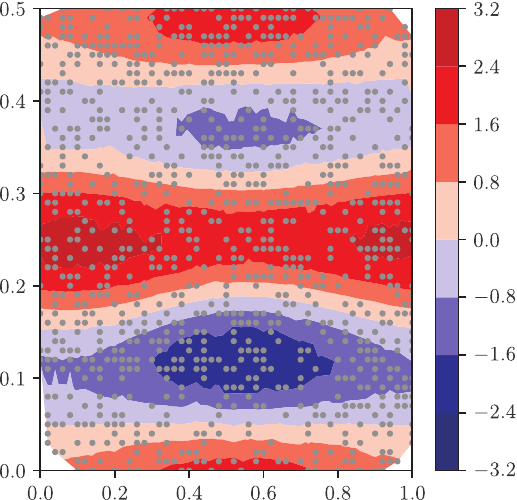}
        \includegraphics[scale=0.39]{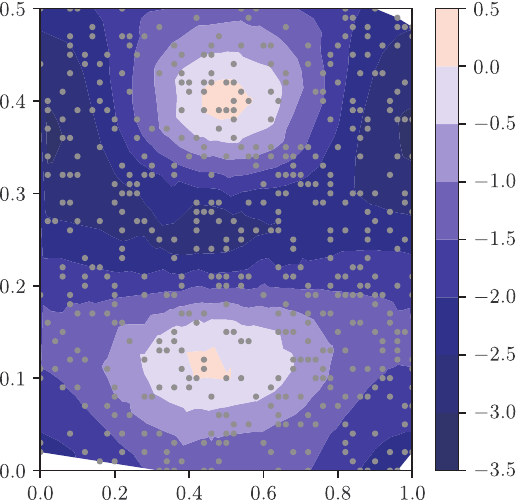}
    }
    \hfill
    \subfloat[Second feature - Curve.]{
        \includegraphics[scale=0.39]{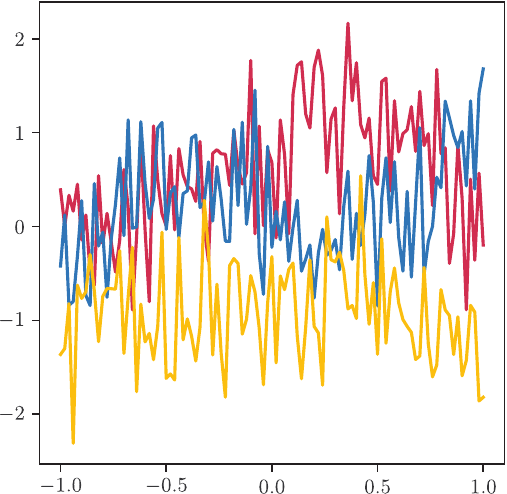}
    }
    \caption{Sample of three observations for the simulated data. For the images, each grey point represents a location where the image is observed. A linear interpolation is performed to plot the images. The colorbars are normalized between $-4$ and $4$.}
    \label{fig:simulation_data}
\end{figure}

To demonstrate the capability of our library in handling a very general class of functional data, we perform a simulation study. The simulation setting is based on \emph{Setting 3} from the simulation in \cite{happMultivariateFunctionalPrincipal2018}. The data generating process uses a truncated version of the Karhunen-Loève decomposition \eqref{eq:kl_multi_trunc} to simulate multivariate functional data with two features ($P = 2$). For the first feature, we generate an orthonormal basis $\{\phi^{(1)}_k\}_{1 \leq k \leq K}$ of $\sLp{}{\TT_{1}}$ on an interval $\TT_{1} = [0, 1] \times [0, 0.5]$ using the tensor product of the first five Fourier basis functions:
\begin{equation}
    \phi^{(1)}_k(s, t) = \psi_l(s) \otimes \psi_m(t), \quad s \in [0, 1] \text{ and } t \in [0, 0.5],\quad k = 1, \dots, K,
\end{equation}
where $\psi_l$ and $\psi_m$ are elements of the Fourier basis. For the second feature, we generate an orthonormal Legendre basis $\{\phi^{(2)}_k\}_{1 \leq k \leq K}$ of $\sLp{}{\TT_{2}}$ on an interval $\TT_{2} = [-1, 1]$. To ensure orthonormality, the bases $\phi^{(1)}_k$ and $\phi^{(2)}_k$ are weighted by random factors $\alpha^{1/2}$ and $(1 - \alpha)^{1/2}$, respectively, where $\alpha \sim \mathcal{U}(0.2, 0.8)$.
Each curve is then simulated using the truncated multivariate Karhunen-Loève expansion \eqref{eq:kl_multi_trunc}:
\begin{equation}
    X(\pointt) = \sum_{k = 1}^K \mathfrak{c}_k \phi_k(\pointt), \quad \pointt \in \TT{} \coloneqq \TT_{1} \times \TT_{2},
\end{equation}
where $\phi_k = (\phi^{(1)}_k, \phi^{(2)}_k)^\top$ and the scores $\mathfrak{c}_k$ are sampled as random normal variables with mean $0$ and variance $\lambda_k$. The eigenvalues $\lambda_k$ decrease exponentially, $\lambda_k = \exp(-(k + 1)/2)$. We simulate $N = 100$ observations. The first feature is sampled on a regular grid of $M^{(1)} = 51 \times 51$ sampling points. We randomly remove between $65\%$ and $85\%$ of the observed points for each observation, resulting in an irregularly sampled functional data. The second feature is sampled on a regular grid of $M^{(2)} = 101$ sampling points. Measurement errors are added to this component using \eqref{eq:model} with $\varepsilon^2 = 0.25$.  We set $K = 25$. The data have been generated using the simulation toolbox provides with the package. Figure~\ref{fig:simulation_data} presents a sample of three observations from the simulated data. 

\begin{figure}
    \centering
    \subfloat[True eigenfunctions for the first feature.]{
        \includegraphics[scale=0.3]{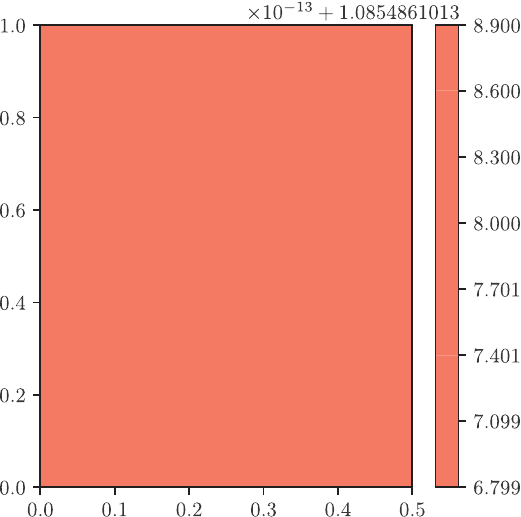}
        \includegraphics[scale=0.3]{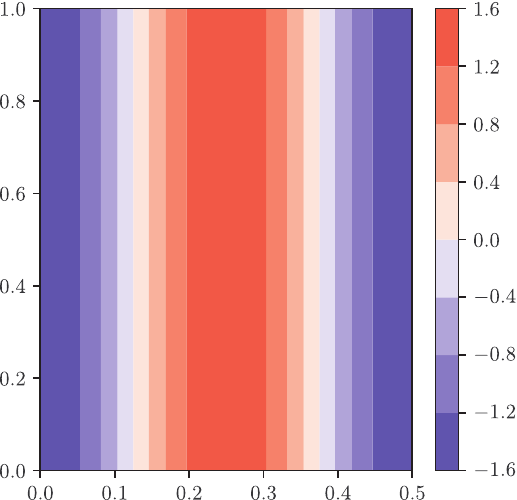}
        \includegraphics[scale=0.3]{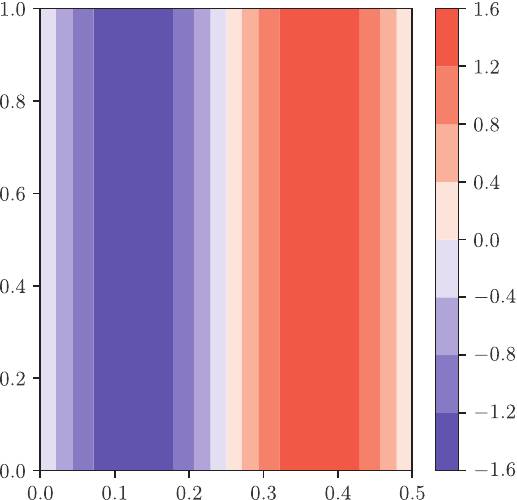}
        \includegraphics[scale=0.3]{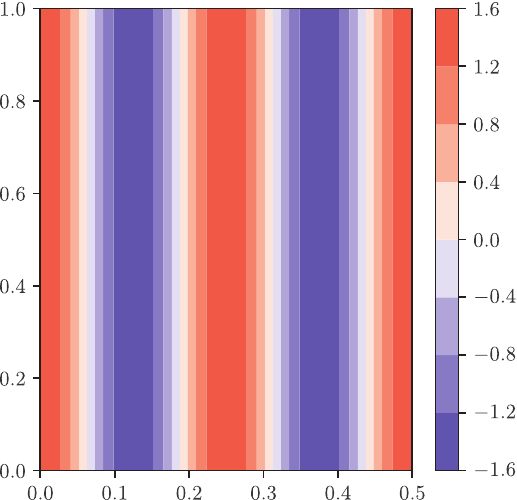}
        \includegraphics[scale=0.3]{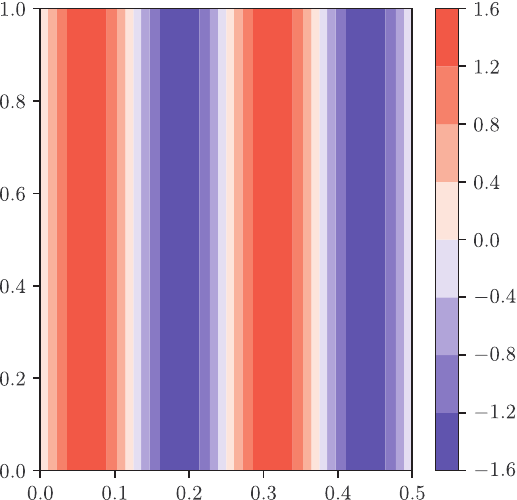}
    }
    \\
    \subfloat[Estimated eigenfunctions for the first feature using MFPCA.]{
        \includegraphics[scale=0.3]{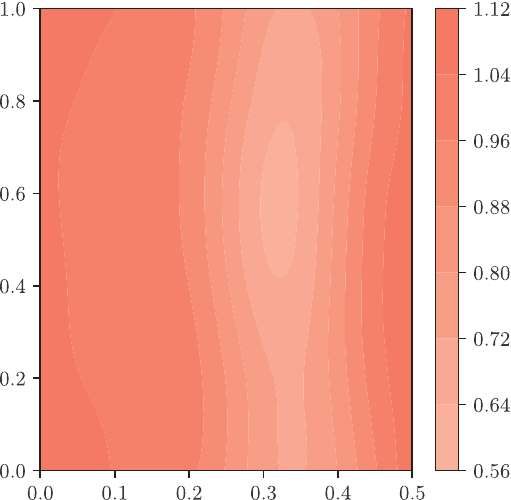}
        \includegraphics[scale=0.3]{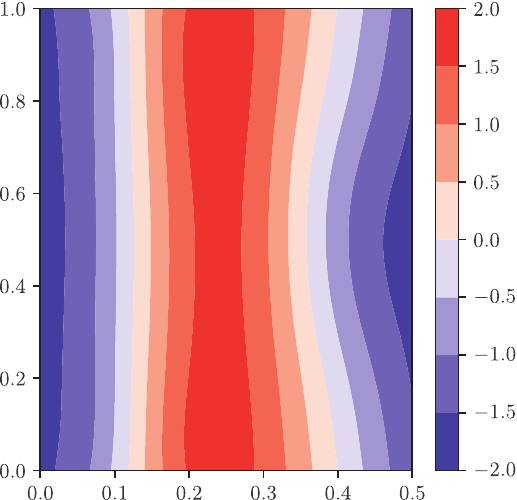}
        \includegraphics[scale=0.3]{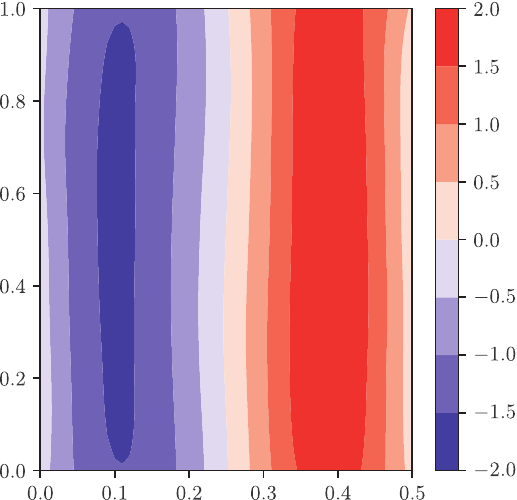}
        \includegraphics[scale=0.3]{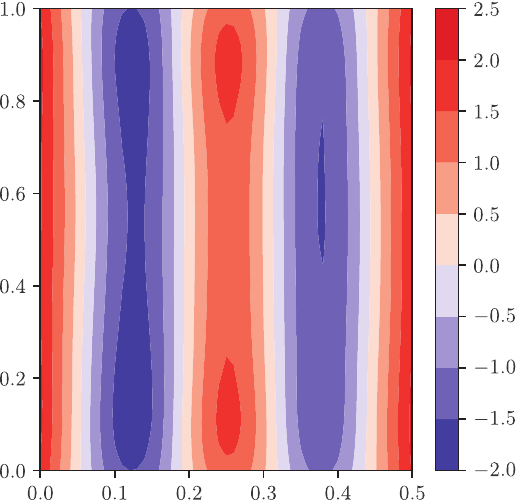}
        \includegraphics[scale=0.3]{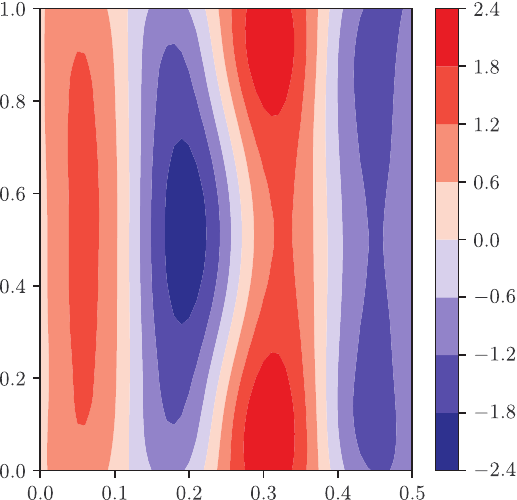}
    }
    \\
    \subfloat[True and estimated eigenfunctions for the second feature.]{\includegraphics[scale=0.6]{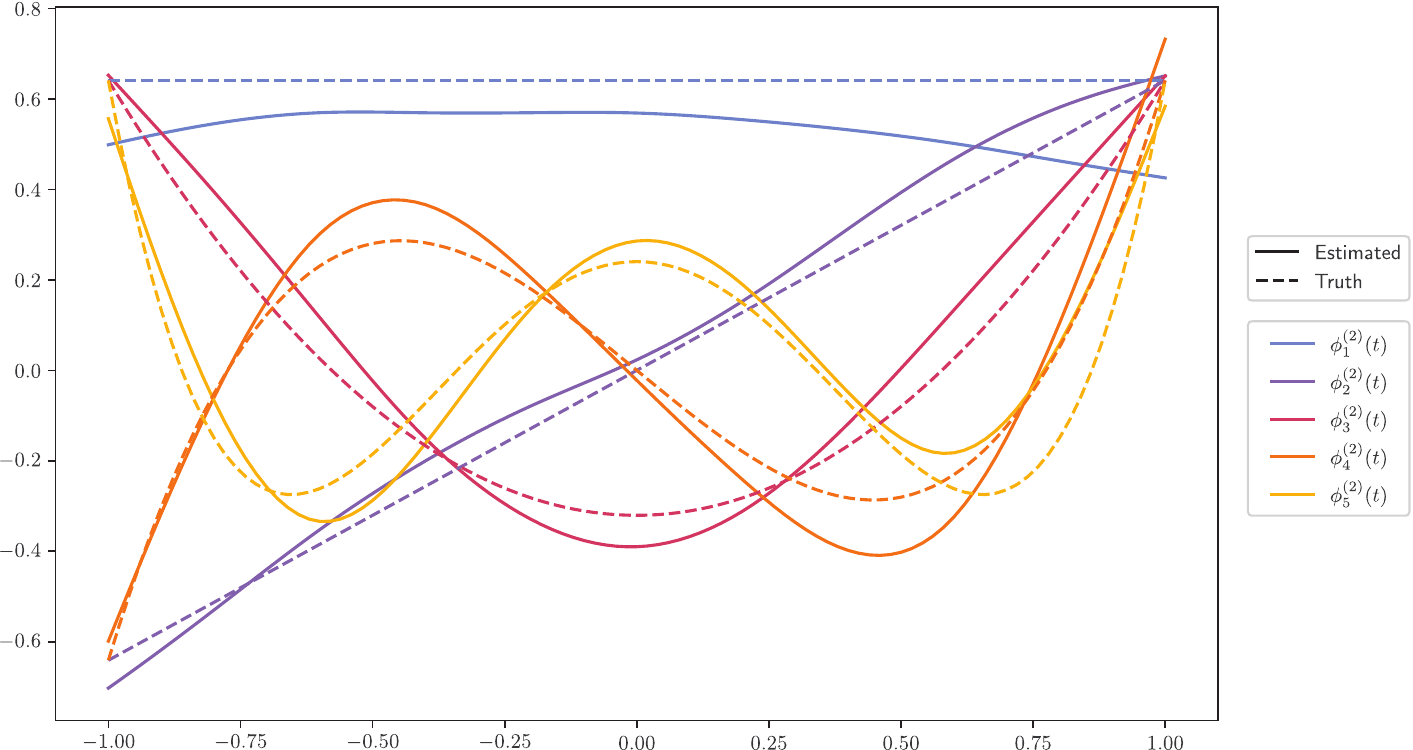}}
    \caption{Comparison between the true and estimated eigenfunctions for the simulated data. For the images, the colorbars are normalized between $-4$ and $4$.}
    \label{fig:simulation_eigenfunctions}
\end{figure}

We aim to estimate the multivariate eigenfunctions. The estimation procedure is run using the inner-product matrix for the first five eigencomponents. The resulting eigenfunctions are smoothed using P-splines smoothing \cite{eilersPracticalSmoothingJoys2021}. Figure~\ref{fig:simulation_eigenfunctions} presents a visual comparison between the true eigenfunctions and their estimation using the diagonalization of the inner-product of the data.

The simulation study demonstrates the capability of our library in processing complex multivariate functional data. The true and estimated eigenfunctions are nearly identical, which highlights the accuracy of the MFPCA implementation and its potential for practical applications in various fields.


\section{Conclusion} 
\label{sec:conclusion}

We propose \textsf{FDApy}, an open-source Python library for the analysis of functional data, including those that are irregularly sampled and/or multivariate. The package provides comprehensive tools for representing functional data using observed values and basis expansion methods. Additionally, a simulation toolbox is included, allowing users to implement and test new methods. The package is publicly available on Github (\url{https://github.com/StevenGolovkine/FDApy}) and the Python Package Index (\url{https://pypi.org/project/FDApy/}). Detailed documentation, including examples, is available at \url{https://fdapy.readthedocs.io/en/latest/}. Tests are implemented using \texttt{unittest}, the unit testing framework provided with Python.

In our examples, \textsf{FDApy} successfully managed functional datasets with diverse characteristics, demonstrating its flexibility and robustness in analyzing multivariate and irregular functional data. The simulation study further highlighted the accuracy of the methods in estimating multivariate eigenfunctions, showcasing its potential for practical applications in various fields.

In the future, we plan to extend the library by implementing regression and classification methods designed for irregular and multivariate functional data. We also encourage others to contribute to the library by adding their methods, following the Contribution Guide in the online documentation. By providing a flexible and robust toolset for functional data analysis, \textsf{FDApy} aims to support researchers and practitioners in uncovering insights from complex functional datasets.


\section*{Acknowledgment}

The author would like to thank Norma Bargary and Andrew J. Simpkin for their insightful comments and careful reading of the manuscript.

\section*{Funding}

This work was supported by Groupe Renault and ANRT (French National Association for Research and Technology) under the CIFRE convention No. 2017/1116; Science Foundation Ireland under Grant No. 19/FFP/7002 and co-funded under the European Regional Development Fund.

\bibliographystyle{abbrvnat}
\bibliography{biblio}

\end{document}